# New example of charge conjugation and parity violation from search for a permanent electric dipole moment of Rubidium atom

Pei-Lin You[1]  Xiang-You Huang [2]

1. Institute of Quantum Electronics, Guangdong Ocean University, zhanjiang 524025, China.
2. Department of Physics, Peking University, Beijing 100871, China.

Quantum mechanics thinks that all atoms do not have permanent electric dipole moment ( EDM )because of their spherical symmetry. Therefore, there is no polar atom in nature except for polar molecules. The electric susceptibility $x_e$ caused by the orientation of polar substances is inversely proportional to the absolute temperature T while the induced susceptibility of atoms is temperature independent. Using special capacitors our experiments discovered that directional motion of Rb atoms in a non-uniform electric field, ground state Rb atom is polar atom with a large EDM: $d(Rb) = [1.70 \pm 0.20 \pm 0.14] \times 10^{-8}$ e.cm. The experiment showed that the relationship between $x_e$ of Rb vapor and T is just $x_e = B/T$, where the slope B≈380(k) as polar molecules. Its capacitance C at different voltage V was measured. The C-V curve shows that the saturation polarization of Rb vapor has be observed when the field $E \geq 8.5 \times 10^4$ V/m. New example of CP (charge conjugation and parity) violation occurred in Rb atoms. If Rb atom has a large EDM, why the linear Stark effect has not been observed? The article discussed the question thoroughly. Our results are easy to be repeated because the details of the experiment are described in the article.

**1.Introduction**   In order for an atom or elementary particle to possess a permanent electric dipole moment (EDM), time reversal (T) symmetry must be violated, and through the CPT theorem CP(charge conjugation and parity) must be violated as well[1]. The currently accepted Standard Model of Particle Physics predicts unobservable the dipole moments of an atom, therefore, EDM experiments are an ideal probe for new physics beyond the Standard Model. Experimental searches for EDMs can be divided into three categories: search for the neutron EDM[2], search for the electron EDM utilizing paramagnetic atoms, the most sensitive of which is done with Tl atoms(the result is $d_e = [1.8 \pm 1.2 \text{ (statistical)} \pm 1.0 \text{ (systematic)}] \times 10^{-27}$ e.cm)[3], and search for an EDM of diamagnetic atoms, the most sensitive of which is done with $^{199}$Hg(the result is $d(Hg) = -[1.06 \pm 0.49 \text{ (stat)} \pm 0.40 \text{ (syst)}] \times 10^{-28}$ e.cm )[1,4]. Experiments to search for an EDM of atom began many decades ago, no large EDM has yet been found [1-5]. In all experiments, they measured microcosmic Larmor precession frequency of individual particle based on nuclear spin or electron spin. The search for an EDM consists of measuring the precession frequency of the particle in parallel electric and magnetic fields and looking for a change of this frequency when the direction of **E** is reversed relative to **B.** We now submit the article on the similar topic, however, with measuring macroscopic electric susceptibility( $x_e$ ) of Rubidium vapor containing a large number of Rb atoms (the density N>10$^{21}$ m$^{-3}$). This article reported four experimental phenomenon which have not been observed. Our experiments showed that ground state Rb atom is polar atom with a large EDM. On the other hand, some evidence for CP violation beyond the Standard Model comes from Cosmology. Astronomical observations indicate that our Universe is mostly made of matter and contains almost no anti-matter. The first example of CP violation was discovered in 1964, but it has been observed only in the decays of the K$_o$ mesons. After 38 years, the BaBar experiment at Stanford Linear Accelerator Center (SLAC) and the Belle collaboration at the KEK laboratory in Japan announced the second example of CP violation in B mesons. "The results gave clear evidence for CP violation in B mesons. However, the degree of CP violation now confirmed is not enough on its own to account for the matter-antimatter imbalance in the Universe." (SLAC Press Release July 23, 2002).This fact suggests that there must be other ways in which CP symmetry breaks down and rarer processes and more subtle effects must be examined. So EDM experiments are now considered an ideal probe for evidence of new sources of CP violation. If an EDM is found, it will be compelling evidence for the existence of new sources of CP violation[6]. **Our experiments showed that new example of CP violation occurred in Rb atoms.** This finding is a vital clue that an unknown factor is very likely at play in Rb atoms**.** The correctness of our theoretical motivation is supported by such five facts as the following.

①The shift in the energy levels of an atom in an electric field is known as the Stark effect. Normally the effect is quadratic in the field strength, but first excited state of the hydrogen atom exhibits an effect that is linear in the strength. This is due to the degeneracy of the excited state. This result shows that the hydrogen atom (the quantum number n=2 ) has very large EDM, $d_H=3ea_o=1.59\times10^{-8}$e.cm ($a_o$ is Bohr radius)[7,8]. On the one hand, this EDM does not depend on the field intensity, hence it is not induce by the external field but is inherent behavior of the atom[7,8]. L.I. Schiff once stated that "Unperturbed degenerate states of opposite parities, as in the case of the hydrogen atom, can give rise to a permanent electric dipole moment"[7]. L.D. Landay also once stated that "The presence of the linear effect means that, in the unperturbed state, the hydrogen atom has a dipole moment"[8]. But on the other hand, the calculation of quantum mechanics tells us that unperturbed degenerate states of hydrogen atom(n=2) with zero EDM and has result $\langle\psi_{2lm}|e\mathbf{r}|\psi_{2lm}\rangle=0$, where $\psi_{2lm}$ are four wave functions of unperturbed degenerate states[8]. Due to the EDM of the hydrogen atom is responsible for the presence of linear Stark effect. A hydrogen atom (n=2) with zero EDM how responds to the external field and results in the linear Stark effect ? Quantum mechanics can not answer the problem[9-11]!

②In addition, the radius of the hydrogen atom of the first excited state is $r_H = 4a_o = 2.12\times10^{-8}$ cm, it is almost the same as the radius of $^{199}$Hg ($r_{Hg}=1.51\times10^{-8}$ cm)[12], but the discrepancy between their EDM is by some twenty orders of magnitude! How do explain this inconceivable discrepancy? The existing theory can not answer the problem! The existing theory thinks that in quantum mechanics there is no such concept as the path of an electron[8]. No one will give you any deeper explan of the inconceivable discrepancy. No one has found any more basic mechanism from which these results can be deduced! However, a hydrogen atom (n=2) has a nonzero EDM in the semi-classical theory of atom. The electron in a hydrogen atom (n=2) moves along a quantization elliptic orbit. We can draw a straight line perpendicular to the major axis of the elliptic orbit through the nucleus in the orbital plane. The straight line divides the elliptic orbit into two parts that are different in size. According the law of conservation of angular momentum, the average distance between the moving electron and the static nucleus is larger and the electron remains in the large part longer than in the small part. As a result, the time-averaged value of the electric dipole moment over a period is nonzero for the atom.

③The alkali atoms having only one valence electron in the outermost shell can be described as hydrogen-like atoms[13]. Since the quantum number of the ground state alkali atoms are n≥2 rather than n=1, in the Sommerfeld picture the valence electron moves along a highly elliptical orbit, the so-called diving orbits, approach the nucleus, as the excited state of the hydrogen atom. So we conjecture that the ground state neutral alkali atoms may have large EDM of the order of $ea_o$, but the actual result of the conjecture still needs to be tested by experiments[14].

④Quantum mechanics thinks that atoms do not have EDM because of their spherical symmetry. Therefore, there is no polar atom in nature. When atoms are placed in an electric field, they become polarized, acquiring *induced* electric dipole moments in the direction of the field. On the other hand, many molecules do have EDM. This molecule is called a polar molecule, such as $H_2O$, HCl, etc. When a field is applied, the polar molecules tend to orient in the direction of the field. Note that the susceptibility($x_e$) caused by the orientation of polar molecules is inversely proportional to the temperature(T): $x_e = B/T$ while the induced susceptibility due to the distortion of electronic motion in atoms is temperature independent: $x_e = A$, where A and the slope B is constant, $x_e = C/C_o - 1$, $C_o$ is the vacuum capacitance and C is the capacitance of the capacitor filled with the material[15]. J.D. Jackson once stated that this difference in temperature dependence offers a means of separating the polar and non-polar substances experimentally [15]. The molecular electric dipole moments are of the order of magnitude of the electronic charge ($1.6\times10^{-19}$ C) multiplied by the molecular dimensions ($10^{-10}$ m), or about $10^{-30}$ coulomb-meters.

⑤R.P. Feynman considered the orientation polarization of water vapor [16]. He plotted the straight line from four experimental points $x_e = B/T$, where $T_1=393$K and $x_{e1}=400.2\times10^{-5}$, $T_2=423$K and $x_{e2}=371.7\times10^{-5}$, $T_3=453$K and $x_{e3}= 348.8\times10^{-5}$, $T_4=483$K and $x_{e4}=328.7\times10^{-5}$[17]. We work out the slope B≈1.50 K, where $x_e = Nd_o^2/3kT\varepsilon_o$ and $B=Nd_o^2/3k\varepsilon_o$. Notice that k is Boltzmann constant, $\varepsilon_o$ is the permittivity of free space, N is the number density of molecules, $d_o$ is the EDM of a molecule[16]. From the slope B≈1.50 K the EDM of a water molecule can be deduced to be $d_{H_2O}= (3k\varepsilon_oB/N)^{1/2}=6.20\times10^{-30}$ C. m≈$0.73ea_o$ and from B≈0.94 K the EDM of a HCl molecule can be deduced to be $d_{HCl} = 3.60\times10^{-30}$ C. m≈$0.43ea_o$ [18]. **If Rb atom is the polar atom, a temperature dependence of the form $x_e = B/T$ should be expected in measuring the susceptibility $x_e$.**

## 2. Experimental method and result

The longitudinal section of the first experiment apparatus is shown in **Fig 1.** There are two glass cylindrical capacitors which are connected and both are filled with Rb vapor. The inside and outside surfaces of the container are silver plated. The silver plated surfaces $a_1$ and $b_1$ constitutes capacitor $C_{11}$ and another two surfaces $a_2$ and $b_2$ constitutes $C_{22}$, G stands for a glass plate which has a round hole in the center and a small hollow glass pipe, the top is shorter compare with the bottom, comes through this hole. A straight conducting wire D which comes through this pipe and surfaces $a_1a_2$ constitutes two electrodes. Before the electric field is switched on, the Rb vapor on both sides is uniform. There is net force **F** on a dipole in non-uniform field:

$$\mathbf{F}= \mathbf{d_o}\nabla \mathbf{E} \qquad (1)$$

where $\nabla E$ is the field gradient and $\mathbf{d_o}$ is the EDM of the dipole. So the dipole tends to move in the direction in which the field increases. After exerting a direct-current voltage between D and $a_1a_2$, a non-uniform field occurs between $H_1$ and $H_2$, Rb atoms immediately move toward $H_1$ where the field is stronger as dipoles. Therefore, the density of Rb vapor changes together with $C_{11}$ and $C_{22}$. The effect was apparently, only by the naked eye, we can find that the density of Rb vapor difference between both sides was large: on the upper half there are liquid Rb particles condensing, while on the other side glass capacitor turns nearly to be transparent. If a Rb atom has no EDM, the net force on the atom is zero in the non-uniform field, the Rb atoms still point in random directions. This experiment result can be hardly interpreted except for Rb atom is polar atom!

The second experiment: investigation of the relationship between the capacitance of Hg or Rb vapor and the density N. The experimental apparatus were two closed glass containers containing Hg or Rb vapor. A stainless steel tube **a** and a silver layer **b** constitute the glass capacitor( Fig.2 , where the plate area $S_1=(5.8\pm 0.1)\times 10^{-2}$ m$^2$, the separation $H_1= H - \triangle =(8.5\pm 0.1)$mm). The radius of the tube **a** and the layer **b** is shown respectively by r and R. Since R-r=H<<r, the capacitor could be approximately regarded as a parallel-plane capacitor.

This capacitor is equivalently connected in series by two capacitors. One is called C and contains the Hg or Rb vapor of thickness $H_1$, another one is called C'' and contains the glass medium of thickness $\triangle$ =1.2mm. The total capacitance C' is C' = C C''/ (C +C'') or C= C'C''/ (C'' - C'), where C'' and C' can be directly measured. The capacitance was measured by a digital meter. The precision of the meter was 0.1pF, the accuracy was 0.5% and the surveying voltage was V=(1.0$\pm$0.001) volt. It means that the applied field only E=V/$H_1\approx 1.2\times 10^2$v/m is weak using the meter. When the two containers are empty, they are pumped to vacuum pressure P $\leqslant 10^{-6}$ Pa for 20 hours. The measured capacitances are nearly the same: $C_{H0}$=(59.2$\pm$0.1) pF(for Hg) and $C_{R0}$=(60.0$\pm$0.1) pF (for Rb). Next, a small amount of Hg or Rb material with high purity is put in the two containers respectively. The two container are again pumped to P $\leqslant 10^{-6}$ Pa, then they are sealed. Now, their capacitances are $C_H$=(60.8$\pm$0.1)pF(for Hg) and $C_R$=(98.6$\pm$0.1)pF(for Rb) respectively. We put the two capacitors into a temperature-control stove, raise the temperature of the stove very slow and keep at $T_1$=(523$\pm$0.5)K for 3 hours. It means that the readings of capacitance are obtained under the saturated vapor pressure of Hg or Rb. The two comparable curves are shown in **Fig.3**. From Ref.12, we obtain the saturated pressure of Hg vapor $P_H$=9914 Pa at $T_1$=523K and measuring capacitance $C_{Ht}$=(63.6$\pm$0.2) pF( $x_e$= $C_{Ht}/C_{H0}$ - 1$\approx$0.07<<1). The formula of saturated pressure of Rb vapor is **P=10$^{6.976 - 3969.5/T}$** psi, the effective range of the formula is 523K $\leqslant$T $\leqslant$643K [12]. We obtain the saturated pressure of Rb vapor $P_R$=0.243 psi =1677Pa at 523K and measuring capacitance $C_{Rt}$= (6550$\pm$10)pF. From the ideal gas law N = P/kT, the density of Rb vapor **$N_1$ = P/k $T_1$ =2.32$\times 10^{23}$ m$^{-3}$**. The experiment shows that the number density $N_H$ of Hg atom is 5.91 times as $N_R$ of Rb atom but the capacitance change ($C_{Ht}$ - $C_{H0}$) of Hg vapor being only 1/1475 of ($C_{Rt}$ - $C_{R0}$) of Rb vapor!

The third experiment: investigation of the relationship between the electric susceptibility($x_e$) of Rb vapor and temperatures(T) at a fixed density $N_2$. The apparatus was the same as the second experiment but at a fixed density $N_2$. The capacitance C of the capacitor was still measured by the digital meter, and its vacuum capacitance $C_{20}$=(70.0$\pm$0.2)pF (where $S_2$= (5.4$\pm$0.1)$\times 10^{-2}$ m$^2$ , $H_2$=(6.8$\pm$0.1)mm ). After the capacitor filled with Rb vapor we put it into the stove and raise the temperature of the stove slow. The capacitance has been measured at several different temperatures T, chosen such that the density $N_2$ of Rb vapor remained fixed. The experimental results are shown in **Fig.4**. We obtain $x_e$ =A+B/T$\approx$B/T, where the intercept A at 1/T= 0 represents the susceptibility due to induced polarization at this density and A$\approx$0, while the slope B=(380$\pm$6)K.

The fourth experiment: measuring the capacitance of Rb vapor at various voltages (V) under a fixed $N_2$ and $T_2$. The apparatus was the same as the third experiment($C_{20}$=(70.0$\pm$0.2)pF) and the method is shown in **Fig.5**. C was the capacitor filled with Rb vapor to be measured, which was kept at $T_2$=(373$\pm$0.5)K., $C_d$ was used as a standard capacitor. Two signals $V_c(t)=V_{co}\cos \omega t$ and $V_s(t)=V_{so}\cos \omega t$ were measured by a digital oscilloscope having two lines. The two signals had the same frequency and always the same phase at different voltages when the frequency is higher than a certain value. It indicates that capacitor C filled with Rb vapor was the same as $C_d$, a pure capacitor without loss. From **Fig.5**, we have C= $C_d$ $(V_{so}-V_{co})/V_{co}$. In the experiment $V_{so}$ can be adjusted from zero to 800V. The capacitance C of Rb vapor at various voltages is shown in **Fig.6.** When $V_{co}=V_1\leq 0.4V$, $C_1\approx$ 142.0pF is approximately constant, $x_e=C_1/C_{20}-1\approx 1.029$ and the Rb vapor is polar substance. With increase of voltage, the capacitance decreases gradually. When $V_{co}=V_2\geq(580\pm 5)V$, $C_2\leq(74.0\pm 0.2)pF$ it approaches saturation, $x_e=C_2/C_{20}-1=0.0571$ and the Rb vapor is non-polar substance. If all the dipoles in a gas turn toward the direction of the field, this effect is called the saturation polarization. The C-V curve shows that the saturation polarization of Rb vapor is obvious when $E=V_2/H_2\geq 8.5\times 10^4 V/m$.

### 3. Discussion

①Our experimental apparatus is a closed glass container. It resembles a Dewar bottle flask in shape. In all experiments, when the capacitor is empty, it is pumped to vacuum pressure $P\leq 10^{-6}$ Pa for 20 hours. The aim of the operation is to remove carefully impurities, such as oxygen, absorbed on the inner walls of the container. In addition, the purity of Rb material exceeded 99.95%. **Actual result showed that the Rb vapor filled the capacitors is present in atomic form, not the Rb dimer.**

②The electric susceptibility of polar molecules is of the order of $10^{-3}$ for gases at NTP and $10^0$ (or one) for solids or liquid at room temperature [15]. Experimentally, typical values of the susceptibility are 0.0046 for HCl gas, 0.0126 for water vapor, 5-10 for glass, 3-6 for mica[18]. When the surveying voltage V=1.0 volt using the digital meter, our experiments showed that the susceptibility of Rb vapor $x_e=C_R/C_{R0}-1=98.6/60.0-1>0.6$ ( where the density $N\approx 10^{21}m^{-3}$), and $x_e=C_{Rt}/C_{R0}-1=6550/60.0-1>100$ (where $N=2.32\times 10^{23}$ $m^{-3}$)! **Few experiments in atomic physics have produced a result as surprising as this one.**

③The electric susceptibility caused by the orientation of gaseous polar molecules is[19]

$$x_e = N\, d_o\, L(a)/\varepsilon_o\, E \qquad (2)$$

where $\varepsilon_o$ is the permittivity of free space, N is the number density of molecules, $d_o$ is the EDM of a molecule. The mean value of $\cos\theta$: $<\cos\theta> = \mu\int_0^\pi \cos\theta \exp(d_oE\cos\theta/kT)\sin\theta\, d\theta = L(a)$, where $a = d_oE/kT$, $\mu$ is a normalized constant, $\theta$ is the angle between $\mathbf{d_o}$ and $\mathbf{E}$. $L(a) = [(e^a+e^{-a})/(e^a-e^{-a})]-1/a$ is called the Langevin function, where a<<1, $L(a)\approx a/3$ and a>>1, $L(a)\approx 1$[19]. When the saturation polarization appeared, $L(a)\approx 1$ and this will happen only if a>>1[19]. R.P. Feynman once stated that *" when a filed is applied, if all the dipoles in a gas were to line up, there would be a very large polarization, but that does not happen"* [16]. So, no scientist has observed the saturation polarization effect of any gaseous dielectric till now! **The saturation polarization of Rb vapor in ordinary temperatures is an entirely unexpected discovery.**

④Due to the induced dipole moment of Rb atom is $d_{int} = G\varepsilon_o E$, where $G=47.3\times 10^{-30}m^3$ [20], the most field strength is $E_{max}\leq 8.5\times 10^4 v/m$ in the experiment, then $d_{int}\leq 3.6\times 10^{-35}$ C.m can be neglected. We obtain

$$x_e = N\, d\, L(a)/\varepsilon_o E \qquad (3)$$

where d is the EDM of an Rb atom and N is the number density of Rb atoms. **L(a)= <$\cos\theta$> is the ratio of Rb atoms which lined up with the field in the total atoms.** Note that $x_e = \varepsilon_r-1 = C/C_o-1$, where $\varepsilon_r$ is the dielectric constant, $E=V/H$ and $\varepsilon_o= C_o H/S$, leading to

$$C - C_o = \beta\, L(a)/a, \qquad (4)$$

**This is the polarization equation of Rb atoms**, where $\beta = SNd^2/kTH$ is a constant. Due to $a=dE/kT= dV/kTH$ **we obtain the formula of atomic EDM**

$$d_{atom} = (C-C_o)V/L(a)SN \qquad (5)$$

In order to work out **L(a)** and **a** of the second experiment, note that in the fourth experiment when the field was weak ($V_1=0.4$ v), $a_1<<1$ and $L(a_1)\approx a_1/3$, $C_1-C_o = \beta/3$ and $\beta=216$ pF. When the field was strong($V_2= 580$ v), $a_2>>1$ and $L(a_2)\approx 1$, $C_2-C_o = L(a_2)\beta/a_2\approx \beta/a_2$. When $a_2\to\infty$, $x_e=0$ and the dielectric constant of the Rb vapor was the same as vacuum! We work out $a_2\approx \beta/(C_2-C_o)=54$, $L(a_2)\approx L(54)=0.9815$. Due to $\mathbf{a}=dE/kT=dV/kTH$, so $\mathbf{a/a_2}=VT_2H_2/T_1H_1V_2$, $a=0.05312<<1$, $L(a)\approx a/3=0.0177$. $L(a)\approx 0.0177$ shows that only 1.77% of Rb atoms have be oriented in the direction of the field when the surveying voltage V=1.0volt or $E=V/H_1\approx 1.2\times 10^2 V/m$. Substituting the values, L(a), $S_1$, $N_1$ and $C-C_o = C_{Rt}-C_{R0} = (6490\pm 10)$pF, we obtain

$$d(Rb)=(C-C_o)V/L(a)S_1N_1 = 2.72\times 10^{-29} C.m = 1.70\times 10^{-8} e.cm \approx 3.2\, ea_o \qquad (6)$$

The occasional errors about the measured value is $\triangle d_1/d=\triangle C/C+\triangle C_o/C_o+\triangle S_1/S_1 +\triangle V/V+\triangle N_1/N_1<0.12$, considering all sources of systematic error $\triangle d_2/d<0.08$, and the combination error $\triangle d/d<0.15$. We obtain

$$d(Rb)=[2.72\pm 0.33(stat)\pm 0.22(syst)]\times 10^{-29} C.m = [1.70\pm 0.20(stat)\pm 0.14(syst)]\times 10^{-8} e.\,cm \qquad (7)$$

**Although above calculation is simple, but no physicist completed the calculation up to now!**

⑤The formula $d_{atom}=(C-C_o)V/L(a)SN$ can be justified easily. The magnitude of the dipole moment of an Rb atom is d = e r. N is the number of Rb atoms per unit volume. L(a) is the ratio of Rb atoms lined up with the field in the total number. Suppose that the plates of the capacitor have an area S and separated by a distance H, the volume of the capacitor is SH. When an electric field is applied, the Rb atoms tend to orient in the direction of the field as dipoles. On the one hand, the change of the charge of the capacitor is $\triangle Q=(C-C_0)V$. On the other hand, when the Rb atoms are polarized by the orientation, the total number of Rb atoms lined up with the field is SHNL(a). The number of layers of Rb atoms which lined up with the filed is H/r. Because inside the Rb vapor the positive and negative charges cancel out each other, the polarization only gives rise to a net positive charge on one side of the capacitor and a net negative charge on the opposite side. Then the change $\triangle Q$ of the charge of the capacitor is $\triangle Q=$SHN L(a)e / (H/r)= SN L(a)d = $(C-C_0)V$, so the EDM of an Rb atom is **d = (C − $C_0$ )V/ SN L(a).**

⑥If Rb atom has a large EDM, why has not been observed in other experiments? This is an interesting question. In Eq.(4) let the function f (a)= L(a)/a and from f″(a)=0, we work out the knee of the function L(a)/a at $a_k$=1.9296813≈1.93. Corresponding knee voltage $V_k=V_2 a_k/a_2 \approx$20.7v, knee field $E_k=V_k/H_2 \approx 3.0\times 10^3$v/m. By contrast with the curve in Fig.6, it is clear that our polarization equation is valid. The fourth experiment showed that the saturation polarization of Rb vapor is easily observed. When the saturation polarization of Rb vapor occurred (V≥580 volt), nearly all Rb atoms (more than 98%) are lined up with the field, and C≈$C_o$ ($C_o$ is the vacuum capacitance)! So only under the very weak field (E<$E_k$ i.e. E<$3.0\times 10^3$V/m), the large EDM of Rb atom can be observed. **Regrettably, nearly all scientists in this field disregard the very important problem.**

⑦As a concrete example, let us treat the linear Stark shifts of the hydrogen( n=2) and Rb atom. Notice that the fine structure of the hydrogen (n=2) is only 0.33 cm$^{-1}$, and therefore the fine structure is difficult to observe [21]. The linear Stark shifts of the energy levels is proportional to the field strength: $\triangle$W= d.E=3e$a_o$E=$1.59\times 10^{-8}$ E e.cm. When E=$10^5$V/cm, $\triangle$W=$1.59\times 10^{-3}$ eV, this corresponds to a wavenumber of 12.8 cm$^{-1}$. For the Hα lines of the Balmer series($\lambda$ = 656.3 nm, and the splitting is only 0.014 nm can be neglected)[21], the linear Stark shifts is $\triangle\lambda = \triangle$W$\lambda^2$/hc =0.55 nm. It is so large, in fact, that the Stark shift of the lines is easily observed [21]. Because the most field strength for Rb atoms is $E_{max}=8.5\times 10^4$V/m≈$9.0\times 10^2$V/cm, if Rb atom has a large EDM d=3.2e$a_o$=$1.70\times 10^{-8}$ e.cm, and the most splitting of the energy levels of Rb atoms $\triangle W_{max}$= d.$E_{max}$= 3.2e$a_o E_{max}$ =$1.5\times 10^{-5}$ eV. This corresponds to a wavenumber of $12.3\times 10^{-2}$ cm$^{-1}$. On the other hand, the observed values for a line pair of the first primary series of Rb atom( Z=37, n=5) are $\lambda_1$=794.76 nm and $\lambda_2$=780.02 nm[12]. The magnitude of the linear Stark shift of Rb atoms is about $\triangle\lambda = \triangle W (\lambda_1 + \lambda_2)^2 / 4hc$ = 0.0076nm. **It is so small, in fact, that a direct observation of the linear Stark shifts of Rb atom is not possible.**

The striking feature of our experiments is the data are reliable and the measuring process can be easily repeated in any laboratory. Our experimental apparatus are still kept, we welcome anyone who is interested in the experiments to visit and examine it. A detail explanation of how fill with Rb or Cs vapor at a fixed density in a vacuum environment as indicated in Ref. 22.

**References**
1. M. V. Romalis, W .C. Griffith, J.P. Jacobs and E N. Fortson, Phys. Rev. Lett. **86,** 2505 (2001)
2. P. G. Harris *et al.*, Phys. Rev. Lett. **82,** 904 (1999)
3. E. D. Commins, S. B. Ross, D. DcMille, and B.C. Regan, Phys .Rev. A **50,** 2960 (1994)
4. J. P. Jacobs, W. M. Klipstein, S. K. Lamoreaux, B. R. Heckel, and E. N. Fortson, Phys. Rev. A **52,** 3521 (1995)
5. Cheng Chin, Veronique Leiber, Vladan Vuletic, Andrew J. Kerman, and Steven Chu , Phys. Rev. A **63**, 033401(2001)
6. H. R. Quinn, and M. S. Witherell, Sci. Am., Oct. 1998, PP76-81
7. L.I. Schiff , Quantum Mechanics (New York McGraw-Hill Book Company 1968)P488, PP252-255
8. L.D.Landay, and E.M.Lifshitz, Quantum Mechanics(Non-relativistic Theory) (Beijing World Publishing Corporation 1999)P290,P2
9. L. E. Ballentine, Quantum Mechanics a modern development (World Scientific Publishing Co, Pte Ltd 1998)P286,P373
10. L.E.Ballentine, Yumin Yang ,and J.P.Zibin, Phys. Rev. A. **50** 2854 (1994)
11. Xiang-You Huang, Phys. Lett. A **121** 54 (1987)
12. J.A. Dean, Lange's Handbook of Chemistry (New York: McGraw-Hill ,Inc 1998)Table 4.6,5.3,5.8 and 7.19
13. W. Greiner, Quantum Mechanics an introduction(Springer-verlag Berlin/Heidelberg 1994)PP254-258, P213.
14. Pei-Lin You, Study on occurrence of induced and permanent electric dipole moment of a single atom, J. Zhanjiang Ocean Univ. Vol.20 No.4, 60 (2000) (in Chinese)
15. J.D. Jackson, Classical Electrodynamics (John Wiley & Sons Inc. Third Edition1999), PP162-165, P173
16. R.P. Feynman, R.B. Leighton, and M. Sands, The Feynman lectures on physics Vol.2( Addison-Wesley Publishing Co. 1964), P11-5,P11-3.
17. Sänger, Steiger, and Gächter, Helvetica Physica Acta 5, 200(1932).
18. I.S. Grant, and W.R. Phillips, Electromagnetism. (John Wiley & Sons Ltd. 1975), PP57-65
19. C.J.F. Bottcher, Theory of Electric Polarization (Amsterdam: Elsevier 1973), P161
20. D.R. Lide, Handbook of Chemistry and Physics. (Boca Raton New York: CRC Press 1998), 10.201-10.202
21. H.Haken, and H.C.Wolf, The Physics of Atoms and Quanta. (Springer-Verlag Berlin Heidelberg 2000),P195,PP171-259
22. Pei-Lin You and Xiang-You Huang , arXiv.0809.4767.

**Acnowledgement**    The authors thank to our colleagues Zhao Tang , Rui-Hua Zhou, Zhen-Hua Guo, Ming- jun Zheng, Xue-ming Yi, , Xing Huang, and Engineer Jia You for their help in the work.

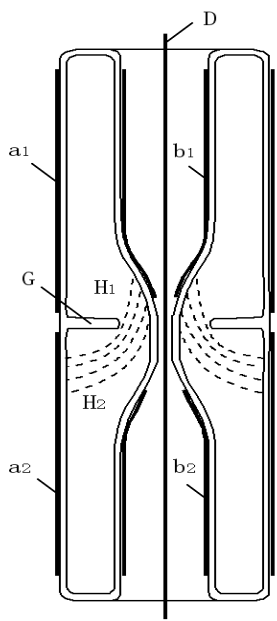

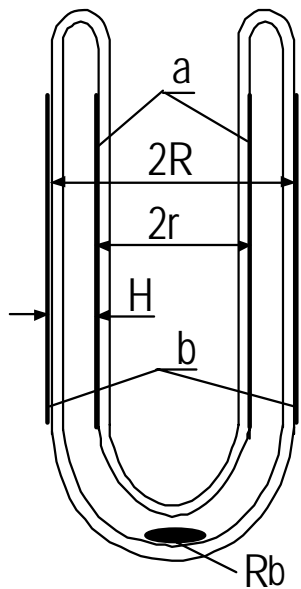

**Fig.1** This is the longitudinal section of the first experimental apparatus. A non-uniform field occurs between $H_1$ and $H_2$, and Rb atom moves toward $H_1$ where the field is stronger. The dashed shows the field lines(not to scale).

**Fig.2** This is the longitudinal section of another three experimental apparatus. Since R-r=H<<r, the capacitor could be regarded as a parallel-plane capacitor.

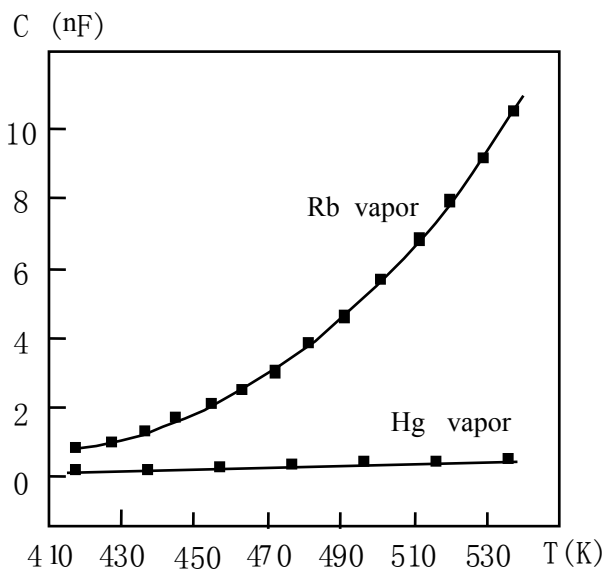

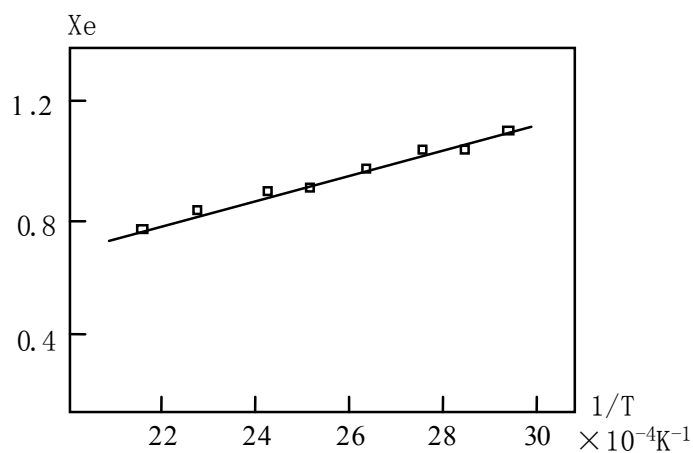

**Fig.3** Two experimental curves showed that the relationship between the capacitance C of Rb or Hg vapor and the density N respectively under saturated vapor pressure, where $1nF=10^3 pF$.

**Fig.4** The experiment showed that the relationship between $x_e$ of Rb vapor and T at a fixed density is just $x_e=A+B/T \approx B/T$, where the slope B≈380K as polar molecules.

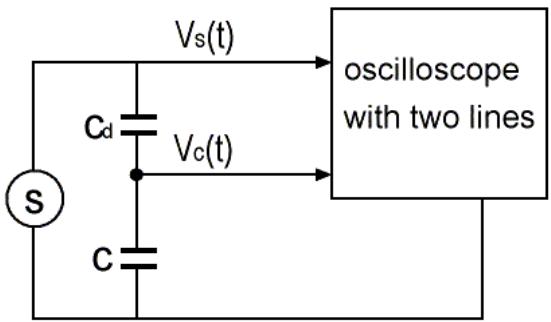
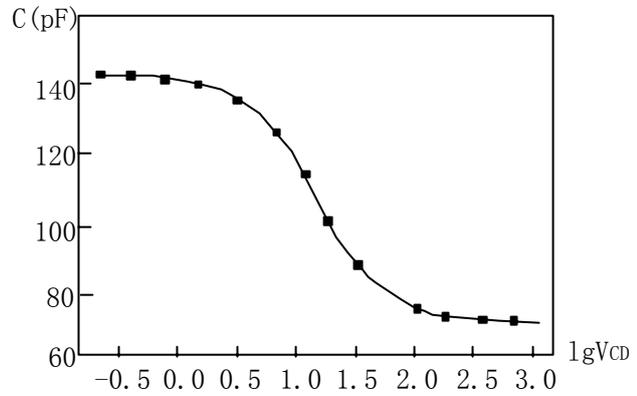

**Fig.5** The diagram shows the experimental method, in which C is capacitor filled with Rb vapor to be measured and Cd is a standard one, where $V_s(t) = V_{so} \cos \omega t$ and $V_c(t) = V_{co} \cos \omega t$.

**Fig.6** The experimental curve shows that the saturation polarization of the Rb vapor is obvious when $E \geqslant 8.5 \times 10^4$ V/m.